# Kossel interferences of proton-induced x-ray emission lines in periodic multilayers

Meiyi Wu (吴梅忆)[1,2], Karine Le Guen[1,2], Jean-Michel André[1,2], Vita Ilakovac[1,2,3], Ian Vickridge[4,5], Didier Schmaus[4,5,6], Emrick Briand[4,5], Sébastien Steydli[4], Catherine Burcklen[7], Françoise Bridou[7], Evgueni Meltchakov[7], Sébastien de Rossi[7], Franck Delmotte[7], Philippe Jonnard[1,2]

[1] Sorbonne Universités, UPMC Univ Paris 06, Laboratoire de Chimie Physique-Matière et Rayonnement, 11 rue Pierre et Marie Curie, F-75231 Paris cedex 05, France
[2] CNRS UMR 7614, Laboratoire de Chimie Physique-Matière et Rayonnement, 11 rue Pierre et Marie Curie, F-75231 Paris cedex 05, France
[3] Université de Cergy-Pontoise, F-95031 Cergy-Pontoise, France
[4] Sorbonne Universités, UPMC Univ Paris 06, Institut des NanoSciences de Paris, 4 place Jussieu, boîte courrier 840, F-75252 Paris cedex 05, France
[5] CNRS UMR 7588, Institut des NanoSciences de Paris, 4 place Jussieu, boîte courrier 840, F-75252 PARIS cedex 05, France
[6] Université Paris Diderot-P7, F-75205 Paris cedex 13, France
[7] Laboratoire Charles Fabry, Institut d'Optique Graduate School, CNRS, Université Paris-Saclay, F-91127 Palaiseau Cedex, France

The Kossel interferences generated by characteristic x-ray lines produced inside a periodic multilayer have been observed upon proton irradiation, by submitting a $Cr/B_4C/Sc$ multilayer stack to 2 MeV protons and observing the intensity of the Sc and Cr Kα characteristic emissions as a function of the detection angle. When this angle is close to the Bragg angle corresponding to the emission wavelength and period of the multilayer, an oscillation of the measured intensity is detected. The results are in good agreement with a model based on the reciprocity theorem. The combination of the Kossel measurements and their simulation, will be a useful tool to obtain a good description of the multilayer stack and thus to study nanometer-thick layers and their interfaces.

**Introduction**

The interferences of characteristic x-ray emission produced within a crystal, by the crystal lattice itself was demonstrated experimentally by Walter Kossel in 1935 [1]. These interferences lead to the observation of what are now called Kossel lines, which are analogous to the Kikuchi lines observed in electron diffraction. Since the observation of a characteristic emission requires as a first step the ionization of a core level, different ionizing radiations can be used to observe Kossel lines [2]:

- Electrons from an electron gun [3–6] or a scanning electron microscope [7];
- X-ray photons from an x-ray tube [8–10] or synchrotron radiation [11–14]; this case is analogous to the x-ray standing wave technique [14,15] used to study the interfaces of multilayers [16] or x-ray waveguides [17] as well as superficial thin films [18];
- Rapid charged particles (proton or ion beam) from an accelerator [19–25].

The technique requires a periodic structure to diffract the emitted radiation, thus it has been applied to study crystals and interferential multilayers. However, to the best of our knowledge, Kossel lines have never been observed in multilayers upon particle excitation. It is the aim of this paper to show that it is possible to study multilayers having a nanoscale period upon proton irradiation through the observation of Kossel lines.

**Experimental details**

The multilayer used for this experiment was a $Cr/B_4C/Sc$ periodic multilayer, whose period was repeated 100 times. The samples were deposited by magnetron sputtering onto a silicon substrate. The thickness of the Cr, $B_4C$ and Sc layers are 0.60, 0.20 and 0.92 nm respectively, as deduced from x-ray reflectivity (XRR) measurements in the hard and soft x-ray ranges. Thus the period of the stack is 1.72 nm. Other reflectivity measurements also showed that Cr atoms are present inside the $B_4C$ layers [26]. The thin $B_4C$ barrier layers were introduced to prevent the interdiffusion between Cr and Sc layers and can also improve the thermal stability of the stack [27]. The multilayer was capped with a 2.5 nm-thick $B_4C$ layer in order to prevent it from oxidation. Working with this $Cr/B_4C/Sc$ multilayer enabled us to measure well-resolved Sc and Cr K lines and to detect them at reasonable grazing angles. Indeed, this kind of short-period multilayer is chosen for microscopy or spectroscopy applications in the water window range [28,29] and requires well defined layers.

Protons of 2.0 MeV produced by the Van de Graaff accelerator of the SAFIR Platform of the Université Pierre et Marie Curie were used to excite the sample. With such energy, the protons ionise the Sc and Cr atoms in their K shell uniformly over the full multilayer thickness. The size of the beam on the sample was approximately 2 mm and the beam current maintained between 100 and 150 nA for the duration of the experiments.

Possible evolution of the multilayer composition was monitored simultaneously via the elastically backscattered protons detected in a passivated implanted planar silicon (PIPS) detector, placed at 165° with respect to the direction of the proton beam. We present in Figure 1 the spectrum so obtained for a total proton dose of 600 μC. Protons scattered from the Cr and Sc atoms appear in the peak near 1800 keV. The area of this peak was monitored, and did not change during the measurements, indicating no measurable loss of matter induced by the beam. As a further precaution, the proton beam was moved on the sample surface, from one pristine zone to another, so that the dose did not exceed 600 μC at a given location.

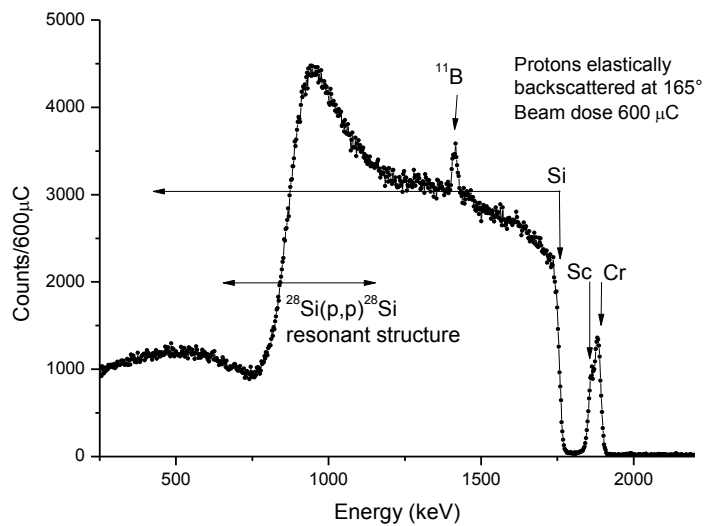

**Figure 1**: 2.0 MeV proton elastic backscattering spectrum obtained from the Cr/B$_4$C/Sc multilayer. The surface energies of the various elements are indicated by the downward arrows. The scattering cross sections on the light elements are highly non-Rutherford: in particular the $^{11}$B(p,p)$^{11}$B cross section is much greater than the Rutherford cross-section which explains why the $^{11}$B peak is visible in spite of the very small quantity of boron, and the highly non-Rutherford $^{28}$Si(p,p)$^{28}$Si cross-section is manifested in the resonant structure obtained from the thick silicon substrate.

The setup for the Kossel experiment is shown in Figure 2. The angle between the incident proton beam and the x-ray detector (SDD, silicon drift detector) was fixed at 90°. In addition to the beryllium window protecting the detector from the atmosphere, a 60 μm thick Mylar film was placed in front of the detector to block the scattered protons. Overall transmission for the Sc K and Cr K lines is greater than 0.65 [30] and practically zero for the Sc and Cr L lines, and also for the B and C K lines.

We rotated the sample so that the angle θ between the sample surface plane and the direction of the SDD, was varied around the Bragg angle, which is around 4° for the Cr Kα emission and 5° for the Sc Kα emission (see below). The distance between the sample and the detector was 115 mm and a slit of ≈0.2 mm width was placed in front of the collimator of the SDD, giving an angular resolution of about 0.1°.

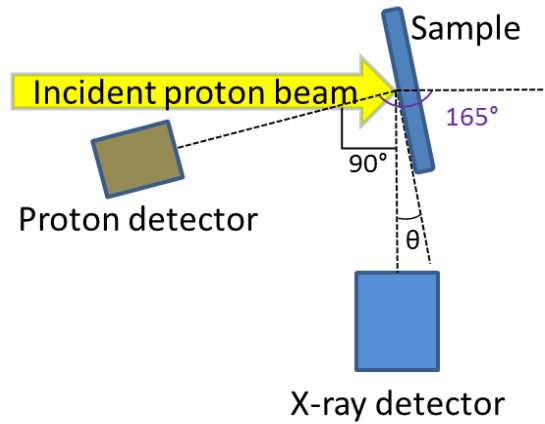

**Figure 2**: Experimental setup for the Kossel experiment. The sample is rotated so that the detection angle θ is around the first order Bragg angle calculated by considering the emission wavelength and multilayer period. The x-ray and proton detectors are fixed.

We show in Figure 3 the proton induced x-ray emission (PIXE) spectrum of the multilayer obtained with a proton dose of 270 μC. The Kα and Kβ lines of both scandium and chromium are well resolved. Their energies are respectively, 4090 and 4460 eV (0.303 nm and 0.278 nm respectively) for scandium and 5414 eV and 5947 eV (0.230 nm and 0.208 nm respectively) for chromium [31]. Moreover, since the proton-induced Bremsstrahlung background is very small, in the following we only consider the intensity under each peak and do not perform any fitting of the spectra or background subtraction. The Kossel experiment consists of observing these intensities as a function of the detection angle θ and plotting them to obtain what we call Kossel curves.

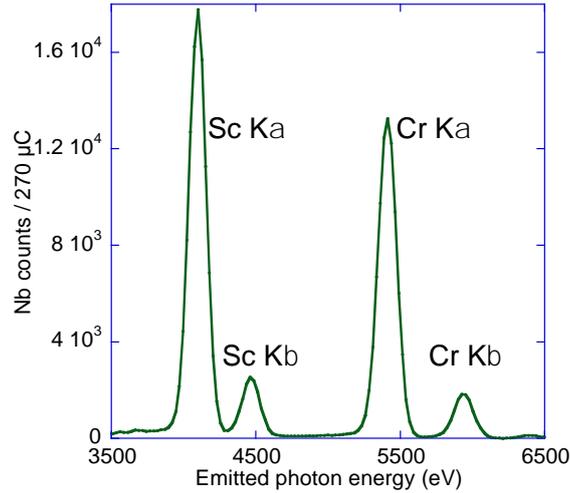

**Figure 3**: X-ray emission spectrum of the Cr/B$_4$C/Sc multilayer induced by 2 MeV protons.

**Results and discussion**

We show in Figure 4 the Kossel curves for the Sc and Cr Kα emissions toward the low angles and starting from zero. At each angle, the proton dose is 30 µC. The angular step is 0.2° except in the region of the strongly varying intensity where it is 0.05°. For both curves, the intensity increases first sharply and then smoothly. The intensity increase is due to the total internal reflection of the radiation emitted within the sample. The angular shift observed in Figure 4, between the onset of the Sc and Cr curves, about 0.15°, comes from the difference of the mean optical indices of the multilayer at the energies of Sc Kα and Cr Kα radiations. The sharp edge inflection point is the angle of the total internal reflection and is chosen to calibrate the angular scale. This last angle is calculated (with a typical uncertainty of ±0.01°), by optical simulation [30] of the reflectance at the two corresponding energies, as 0.56° for Sc and 0.42° for Cr. The uncertainty on the absolute value of the experimental angles is estimated to be 0.05° (the angular step around the inflexion point) while the relative uncertainty is lower than this value.

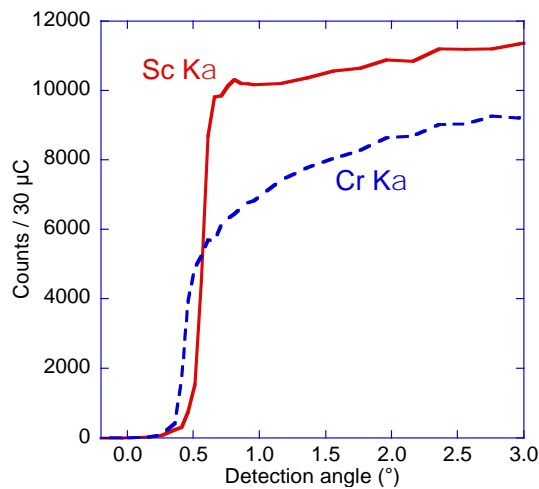

**Figure 4**: Wide scan of the Kossel curves for the Sc Kα (solid curve) and Cr Kα (dashed curve) emissions.

Figure 5 shows the Kossel curves obtained in the range of the Kossel feature with a 0.05° angular step and using the angular calibration determined above. At each angular position, the total dose was 120 μC for the Sc Kα curve, and 150 μC for the Cr Kα curve. The error bars represent one standard error. Experimental curves are compared to simulations [8,9,32] that calculate the electric field generated within the periodic stack by the emitted radiation according to the reciprocity theorem. The parameters of the multilayer (thickness, roughness) introduced in the simulation are those deduced from XRR analyses. The simulations are convolved with an experimental function (a rectangle) having a 0.1° width representing the angular resolution of the experimental setup. Let us note that in the case of proton excitation, simulations are easier to perform than those in the cases of electron and x-ray excitations, where the depth-dependent ionization probability should be considered; electrons loose significant energy within the thickness of the stack while x-rays are either attenuated or even create standing waves within the periodic structure.

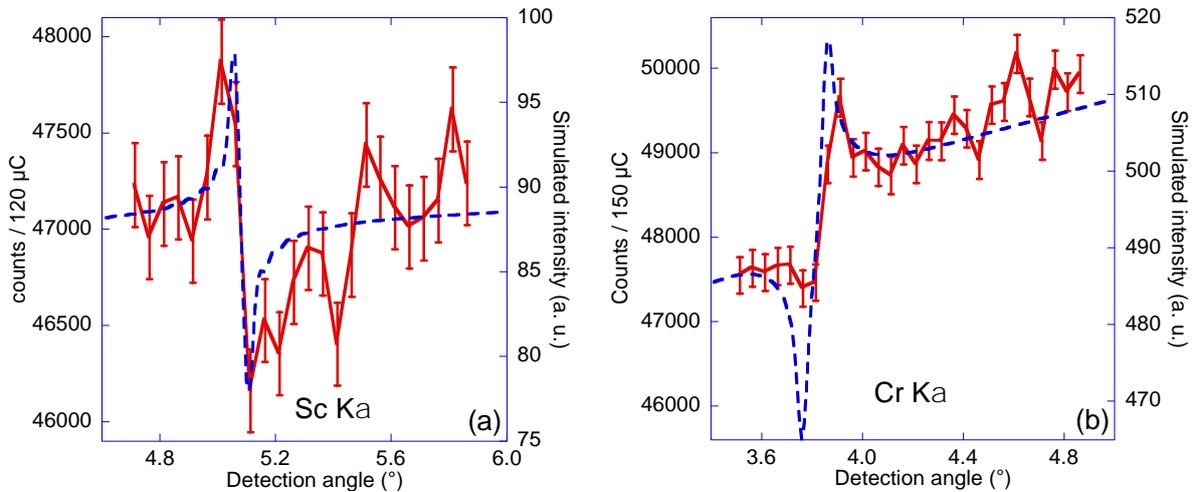

**Figure 5**: Kossel curves for the Sc Kα (a) and Cr Kα (b) emissions displayed in a narrow angular range around the Bragg angle calculated at the first diffraction order with the wavelength of the corresponding radiation and the period of the multilayer: experiments (solid curves); simulations (dashed curves).

The experimental Kossel features, Figure 5, lack a little contrast, *i.e.* the difference of intensity between the maximum and the minimum is small, owing to the relatively large angular resolution and to the limited counting statistics. This last point comes from the required angular aperture of the detection system, which leads to a very small solid angle of collection of the x-rays. It is observed, in agreement with the simulations, that the intensity fluctuations in the case of the Sc Kα curve, passing first by a maximum then by a minimum, separated by about 0.1° (see Fig. 5(a)), are reverse with respect to Cr Kα curve, (see Fig. 5(b)). The inflexion points of the Kossel curve discontinuities, at 5.10° and 3.85° for the Sc and Cr respectively, are consistent (within their ±0.05° uncertainty) with the angular values of 5.086° and 3.840° calculated with Bragg's law corrected from refraction, using the Sc and Cr emission wavelengths and multilayer period. This can be seen from the position of the Kossel features, well in agreement with the simulations.

The intensity modulation observed in the Kossel curve when scanning the detection angle comes from the variation of the location of the electric field inside the stack since changing the angle moves the system of standing waves perpendicular to the layers of the stack. To illustrate this point we consider the Sc Kossel curve and the depth distribution of the electric field, which, around the Bragg angle, has nearly the same period as the multilayer. When the detection angle is equal to the Bragg angle, *i.e.* corresponding approximately to the inflexion point of the sharp intensity decrease, Fig. 5(a), the maxima of the electric field are located at the interfaces between the Sc and $B_4C$ layers. Shifting the detection angle by -0.035° moves the maxima of the electric field to the centre of the Sc layers, as shown as an example in Figure 6. Shifting the detection angle by +0.015° and +0.03° with respect to the Bragg angle moves the maxima of the electric field at the interfaces between the $B_4C$ and Cr layers, and at the centre of the Cr layers, respectively (not shown here). This provides a mean to selectively excite different locations of a multilayer, either inside its layers or at its interfaces.

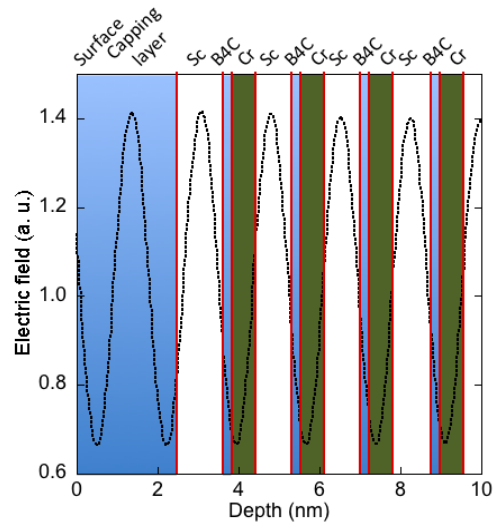

**Figure 6**: Depth distribution of the electric field generated within the $Cr/B_4C/Sc$ multilayer by the emitted Sc Kα radiation. The calculation illustrated here is made for an angle -0.035° away from the Bragg angle. Only the capping layer and the first four periods of the multilayer are represented.

**Conclusion**

In conclusion, we have observed for the first time the Kossel interferences of proton-induced x-ray emission lines in a multilayer. This was achieved by using a Cr/B$_4$C/Sc periodic stack and observing the Sc K$\alpha$ and Cr K$\alpha$ characteristic emissions. Owing to the low background radiation, the x-ray spectra lead directly to the Kossel curves. The agreement between the experiments and the simulation is excellent regarding the shape and position of the Kossel features, although it may be improved regarding the contrast of these features obtained in this pioneering experiment.

The counting statistics for a given incident proton fluence could be improved by changing the angle of incidence between the protons and the sample surface and thus increasing the apparent thickness of the sample seen by the proton beam, or by using a colour camera [33] sensitive both to the location and energy of the detected x-rays. This second solution could provide 1 to 2 orders of magnitude greater x-ray yield per incident proton and could open the way for systematic use of proton-induced x-ray emission with Kossel interferences to probe the nanoscale layers and interfaces of periodic multilayers. Thus, it will become possible to obtain with proton-induced x-ray emission under Kossel interferences the same kind of results and informations as those obtained by the widespread x-ray standing wave technique [34].

*Acknowledgments*: The multilayer sample has been deposited as part of CEMOX (Centrale d'Elaboration et de Métrologie d'Optique X), a platform of LUMAT federation (CNRS FR2764). D. Vernhet and J.-J. Ganem from INSP are thanked respectively for providing us with the x-ray detector and helping during the experiment.